%% file: 9210015.tex

\input paper.tex

\tolerance=1000

\centerline {{\bf COSMOLOGICAL MODELS IN TWO}}
\centerline {{\bf SPACETIME DIMENSIONS}}
\vskip .5 true cm
\centerline {K.C.K. Chan and R.B. Mann}
\vskip .5 true cm
\centerline {{\bf Department of Physics}}
\centerline {{\bf University of Waterloo}}
\centerline {{\bf Waterloo, Ontario }}
\centerline {{\bf Canada, N2L 3G1}}
\vskip 2.5 true cm

\centerline {{\bf Abstract}}

Various physical properties of cosmological models in $(1+1)$ dimensions
are investigated. We demonstrate how a hot big bang and a hot big crunch
can arise in some models. In particular, we examine why particle horizons
do not occur in matter and radiation models. We also discuss under what
circumstances exponential inflation and matter/radiation decoupling can
happen. Finally, without assuming any particular equation of state, we show
that physical singularities can occur in both untilted and tilted
universe models if certain assumptions are satisfied, similar to the
$(3+1)$-dimensional cases.

\par\vfill
\centerline{\hfill WATPHYS TH92/08}
\footline={\hfil}
\eject
\footline={\hss\tenrm\folio\hss}
\pageno=1

\noindent{{\bf 1 Introduction}}

Gravitational theory in two spacetime dimensions continues to provide
theorists with an interesting ``laboratory'' for exploring the foundations of
classical and quantum gravity. The reduction in the number of degrees
of freedom markedly reduces the complexity of a system of $(3+1)$-dimensional
gravity/matter field equations. The resultant simplicity ought to provide an
important understanding of the issues associated with short-distance
problems, black holes, information loss, topology change, singularities and
the cosmological constant problem.

It is well-known that the price one pays for such simplicity is that one is
forced to adopt a new system of field equations in the gravitational
sector, since the $(N+1)$-dimensional Einstein equations are trivial when
$N=1$. A number of approaches [1] have been adopted to address this problem
including the incorporation of extended geometrical structures,
non-locality, constant curvature theories, non-linear curvature
terms, or dilaton couplings.

Perhaps the simplest approach is to set the only measure of curvature
in two dimensions, the Ricci scalar, equal to the trace of the conserved
stress-energy tensor [2]. Specifically,
$$
R=8\pi GT^a{}_a,\qquad  \nabla_aT^a{}_b=0  \quad \eqno(1)
$$
where $R$ is the curvature scalar, $G$ is the $(1+1)$-dimensional
gravitational constant and $T$ is the trace of the energy-momentum
tensor. A number of features single out such a theory for
consideration. The gravity/matter interaction is qualitatively the same as
that in general relativity: stress-energy acts upon spacetime, telling it
how to curve, and spacetime in turn acts upon stress-energy, telling it how
to move. Viewed as a particular type of ``dilaton gravity'', it is the only
such theory in which the dilaton classically decouples from the
gravity/matter system [3]. Furthermore, the system (1) can viewed as the
$D\rightarrow 2$ limit of the $D$-dimensional Einstein equations ($D=N+1$)
provided the $D$-dimensional gravitational constant $G_D$ is appropriately
rescaled by $(D-2)$ [4]. Finally, despite its simplicity, this theory has a
number of remarkable classical and semi-classical features, including a
well-defined Newtonian limit, black holes, a post-Newtonian expansion, FRW
cosmologies, gravitational collapse, and black hole radiation [2,3,5].
These features suggest that the theory is potentially a very useful tool in
studying quantum gravitational and cosmological effects. Since its classical
features are so similar to those of $(3+1)$-dimensional general relativity,
one might hope that its quantization would bear a similar resemblance to
$(3+1)$-dimensional quantum gravity.

To this end, we investigate here the cosmological properties of a $(1+1)$-
dimensional universe based on the gravitational field equation (1). We
begin in section 2 by outlining the basic mathematical and physical
components of the $(1+1)$-dimensional theory based on (1) as pertains to
the study of cosmology. The Ray\-chaud\-huri equation and some of its
solutions, and the Stefan-Boltzmann law ({\it i.e.} the temperature of a
radiation universe) are also obtained. In section 3 we show that
a hot big bang and a hot big crunch arise in an universe with a
certain kind of matter-radiation mixture and a particular equation of
state. In addition, as a consequence of the lower-dimensionality of the
spacetime, particle horizons do not exist in a $(1+1)$-dimensional
Freidmann-Robertson-Walker (FRW) universe in which matter and radiation are
decoupled. In section 4 we investigate decoupling and inflation, showing
under what circumstances such processes arise. The final section concerns
physical singularities of both FRW and tilted universe models for
general equations of state. We show that infinite density singularities
can occur in both models if certain energy conditions are
satisfied. We also give an explicit model in which a
``whimper'' singularity can arise. Conclusions and an appendix concerning
a specific model which can realize inflation round out our work.

Throughout this paper we take speed of light to be unity but retain
Newton's constant $G$ explicitly. We use the metric signature $(-+)$.

\vskip 0.3 true cm

\noindent{\bf 2 FRW Cosmology}

In this section we set up the general formalism suitable for analyzing
$(1+1)$\-dimensional cosmologies.

We shall assume that there exists locally a {well-defined} normalized
velocity two vector $U^a$ $(U^aU_a=-1)$ describing the motion of
the matter in the universe, and a uniquely defined
symmetric one metric
$$
h_{ab}=g_{ab}+U_a U_b, \eqno(2)
$$
which is a local projection tensor $(h^a{}_b h^b{}_c=h^a{}_c)$ projecting
into the one-dimensional instantaneous rest space of
observers moving with $U^a$. This space is
orthogonal to $U^a$, where $h^a{}_a=1$,
$h_{ab}U^a=0$. Thus the spacetime metric $g_{ab}$ induces a spatial
metric $h_{ab}$ on the rest space of $U^a$. As with the
{$(3+1)$-dimensional} case [6], one can write the first covariant
derivative $\nabla_b U_a$ of $U_a$ in $1+1$ dimensions as
$$
\nabla_bU_a=h_a{}^ch_b{}^d\nabla_dU_c-U_bU^d\nabla_dU_a=V_{ab}
-{dU_a\over dt}U_b \eqno(3)
$$
or
$$
\nabla_bU_a=\theta h_{ab}-{dU_a\over dt}U_b,\eqno(4)
$$
where $V_{ab}=h_a{}^c h_b{}^d \nabla_d U_c=\theta h_{ab}$ and
$\theta\equiv\theta^a{}_a$ is the length expansion. The expansion
tensor $\theta_{ab}$ is a symmetric tensor satisfying
$\theta_{ab}U^b=0$; it has only one degree of freedom which may be
expressed by its trace, $\theta$.
The usual vorticity and shear tensors are missing in (4)
because there are no rotations and shears in one spatial dimension.

A spatially homogeneous and isotropic (these two concepts
are equivalent in $1+1$ dimensions, and will be explained in section 5)
model of the universe is a perfect fluid spacetime with the properties:
$$
T_{ab}=ph_{ab}+\rho U_aU_b \eqno(5)
$$
and
$$
{dU^a \over dt}=0. \eqno(6)
$$
Thus the only non-zero kinematic quantity is $\theta$. (6)
implies $U_a=-t_{,a}$, where $t$ measures proper time along all
world lines [6]. Now, the rest spaces defined at each
point by $h_{ab}$ mesh together to form a family of one-dimensional
space-like lines $L_t$ orthogonal to $U^a$. Using
the comoving coordinate $x$ and $t$, we can write the $(1+1)$-dimensional
Friedmann-Robertson-Walker (FRW) metric as
$$
ds^2=-dt^2+{a^2(t)\over 1-kx^2}dx^2,\eqno(7)
$$
where the usual angular coordinates have been suppressed.
A change of variable $dx^2/(1-kx^2) \rightarrow dx^2$ implies
$$
ds^2=-dt^2+a^2(t)dx^2,\eqno(8)
$$
where $a(t)$ is the cosmic scale factor. Thus in $1+1$
dimensions the three different cosmological models
corresponding to values $k=0,\pm 1$ in (7) do not affect
the time evolution of $a(t)$, unlike \hbox{$(3+1)$-dimensional} FRW
universe models. $k=0(-1)$ still describes a spatially open flat
(hyperbolic) universe. $k=1$ describes a closed universe.
In addition, in any FRW model, $a(t)$ is related to the length expansion
$\theta$\ through
$$
{da \over dt} {1 \over a}= \theta. \eqno(9)
$$
For the sections 3 and 4, we shall proceed under the
assumption that the universe is homogeneous and
isotropic.

The Raychaudhuri equation (the time evolution equation
of the expansion $\theta$) in $1+1$ dimensions
can be easily derived. We start
with the Ricci identity applied to the velocity two
vector $U^a$
$$
2\nabla_{[c}\nabla_{d]}U_a=R_{abcd}U^b. \eqno(10)
$$
Multiplying (10) by $U^d$ and then projecting on $a$ and $c$ and using (3),
we get
$$
(h_a{}^ch_b{}^d){dV_{cd}\over dt}-({dU_a\over dt})({dU_b\over dt})
-(h_a{}^ch_b{}^d)\nabla_d({dU_c\over dt})+V_{ad}V^d{}_b+R_{acbd}U^cU^d=0.
\eqno(11)
$$
By contracting $a$ and $b$ in (11), one
finds
$$
{d\theta\over dt}+\theta^2-\nabla_a({dU^a\over dt})+R_{ab}U^aU^b=0\eqno(12)
$$
on using $V_{ab}=\theta h_{ab}$. Using $R_{ab}U^aU^b=4\pi G(\rho-p)$
, which follows from $R_{ab}=(1/2)g_{ab}R$ in
$1+1$ dimensions [7] and (1), for a perfect fluid
and conditions of homogeneity and isotropy $(dU^a/dt=0)$,
the Raychaudhuri equation (12) becomes
$$
{d\theta\over dt}+\theta^2+4\pi G(\rho-p)=0.\eqno(13)
$$
Thus the Raychaudhuri equation in $1+1$ dimensions is quite
similar to the $(3+1)$-dimensional one (see {\it e.g.} [6]) except
for the sign of the pressure term. This difference
prohibits inflation from happening in $1+1$ dimensions
under the influence of scalar fields (as we shall see in section 4).
Furthermore, the term $\rho-p$ plays the role as the
active gravitational mass density and thus the
strong energy condition, SEC, in $1+1$ dimension
becomes $\rho-p>0$. In addition, in section 5,
we shall show that (13) plays an
important role on physical singularities in
\hbox{(1+1)-dimensional} universe models.

For a perfect fluid, we obtain the field equation
$$
{d^2a\over dt^2}=4\pi G(p-\rho)a \eqno(14)
$$
on using (9) and (13). With the comoving velocity
two-vector $U^a=(1,0)$, the conservation law
$$
\nabla^aT_{ab}=0 \eqno(15)
$$
becomes
$$
a{dp\over dt}={d[a(p+\rho)]\over dt}, \qquad or
\qquad {d(\rho a)\over da}=-p \quad .\eqno(16)
$$
This equation describes the
exchange of energy between {matter-radiation} and
gravitational energy.
By considering barotropic perfect fluids
(those where $p=p(\rho)$ being at least $C^1$, so that there is
a \hbox{well-defined} speed of sound $\upsilon_s = (dp/d\rho)^{1/2}$),
we can integrate (16).

Consider the equation of state $p=(\gamma-1)\rho$ with the condition
$2\geq\gamma\geq 1$ as an example. The upper limit corresponds to a
pure radiation universe ($\rho_m=0$), the lower limit to
a pure matter one ($\rho_r=0$), We obtain
$$
p_m=0,\qquad \rho_m\propto a^{-1}\eqno(17)
$$
for a pressure-free matter, and for radiation
$$
p_r=\rho_r,\qquad \rho_r\propto a^{-2}\eqno(18)
$$
on using (16). Both matter and radiation stress-energy tensors
satisfy (16).

In the case of interactions, one may propose a more general situation
$$
\rho_r=\rho_{ro} a^{\beta}, \qquad \rho_m=\rho_{mo} a^{\alpha},\eqno(19)
$$
where $\rho_{ro}$ and $\rho_{mo}$ are non-zero constants in general.
For simplicity, we may assume that the fluids do not interact
significantly,
thereby assuming that both matter and radiation
have almost the same velocity \hbox{two-vector}.
In other words, the total \hbox{energy-momentum} tensor of the fluids is
still in a perfect fluid form with $\rho=\rho_m+\rho_r$ and $p=\rho_r$
(\hbox{pressure-free} matter). Now, substituting (19) into (16), one
obtains
$$
a^{\alpha-\beta}=-{\rho_{ro} \over \rho_{mo}} ({\beta+2 \over
\alpha +1}). \eqno(20)
$$
Note that this equation and the weak energy
condition (WEC) ($\rho_m \geq 0, \rho_r \geq 0 $) impose restrictions on the
values of $\alpha$ and $\beta$. If $a(t)$ is not a constant, we must have
$\alpha=-1$ and $\beta=-2$ (a \hbox{non-interacting} mixture of matter and
radiation) or
$\alpha=\beta$ with $-2<\alpha<-1$.

Now the field equation (14) becomes
$$
{d^2a \over dt^2}=-4\pi G\rho_{mo} a^{\alpha+1}.\eqno(21)
$$
The general solution is
$$
t={1 \over \sqrt{K_1}} \int {{da \over \sqrt{K_2-a^{\alpha+2}}}}
=({a\over \sqrt{K_1 K_2}}){}_2F_1\left({1\over 2},{1\over \alpha+2},
{\alpha+3 \over \alpha+2}; {a^{\alpha+2} \over K_2}\right) \eqno(22)
$$
where $K_1=8\pi G\rho_{mo}/(\alpha+2)$, $K_2$ is an integration constant
and ${}_2F_1$ is the hypergeometric function. This equation seems to imply
that evolution of $a(t)$ depends only on $\alpha$ (or $\rho_m$). However
closer inspection reveals that this is not the case; unless $\beta=-2$ (the
non-interacting case), the value of $\beta$ restricts $\alpha$ via (20) and
the WEC. In this manner the radiation density indirectly affects the
evolution of $a(t)$.

Henceforth, unless otherwise stated, we consider the
\hbox{non-interacting} case. It is clear that the dynamical
equations for the $(1+1)$-dimensional FRW cosmology are
(14), (16) and the equation of state $p=p(\rho)$. For a
mixture of \hbox{non-interacting} matter and radiation universe,
we have
$$
{d^2a\over dt^2}=-4\pi G\rho_{mo} \qquad .\eqno(23)
$$
Thus, such a universe is a dynamical one which began with a
``big bang'' in its
evolutionary history. For a pure radiation universe
({\it i.e.} $\rho_m=0$), $d^2a/dt^2=0$, and so the scale factor
varies linearly with $t$ or it is a constant.
The spacetime is globally equivalent
to flat space since the associated \hbox{energy-momentum} tensor
has vanishing trace and consequently the $(1+1)$-dimensional
curvature tensor vanishes everywhere.

To close this section we obtain the \hbox{Stefan-Boltzmann} law in $(1+1)$
dimensions; this will yield a relation between $a(t)$ and $T$
(temperature) in a radiation universe. Consider, for simplicity,
a one dimensional spatial cavity with length $l$ and whose walls (just two
end-points) are maintained at a temperature $T$. Analogous to the
derivation in three spatial dimensions ({\it e.g.} [8]), one can show that
the number of states of photon in a frequency interval $[f, f+df]$ is
$$
g(f)df=ldf.\eqno(24)
$$
The density states $g(f)$ is therefore constant. From this
and the \hbox{Bose-Einstein} distribution function,
we find that the amount of energy, $dE_f$, carried by the $dn_f$ photons
with frequencies between $f$ and $f+df$ is
$$
dE_f=hfdn_f={lhfdf\over exp[{hf\over kT}]-1}.\eqno(25)
$$
Now, it is easy to see that
the Stefan-Boltzmann law in one spatial dimension is given by
$$
E=\int{dE_f} = lh \int^{\infty}_0 {fdf\over exp[{hf\over kT}]-1}
\eqno(26)
$$
where $E$ is the total energy in the cavity. Rescaling
$(hf/kT)=x$, we see that
$$
{E\over l}=\rho_r\propto T^2.\eqno(27)
$$
Condition (18) then implies
$$
a\propto T^{-1}.\eqno(28)
$$
Thus, in a $(1+1)$ or \hbox{$(3+1)$-dimensional} FRW
universe, the scale factor is inversely proportional to the temperature of
radiation. As $a(t)$ approaches zero, the temperature $T$ becomes infinite
in both cases.

\vskip 0.3 true cm

\noindent{\bf 3 The Big Bang and Particle Horizons}

For a non-interacting mixture of radiation and
matter, one possible solution to (23) is
$$
a(t)=-2\pi G\rho_{mo} t^2+At, \eqno(29a)
$$
where $A=(da/dt)$ at $t=0$. Consider a universe containing such a (non-
interacting) mixture of matter and radiation. As $a(t)$ approaches $0$,
$\rho_r$ has no effect on the spacetime and the time evolution of $a(t)$,
even though $\rho_r$ could be much greater than $\rho_m$. At $t=0$ or
$A/2\pi G\rho_{mo}$, $a(t)\rightarrow 0$ and correspondingly, $\rho$
diverges as does the curvature scalar $R =-8\pi G\rho_m$ and the
temperature $T \propto a^{-1}$. Thus we
have a hot big bang and a hot big crunch in a universe with a
non-interacting mixture of matter and radiation.

Note also that for a pure matter ($\rho_r=0$) universe, one has [5]
$$
a(t)=a_o[1-2\pi G\rho_o (t-{1\over \sqrt{2\pi G\rho_o}})^2], \eqno(29b)
$$
where $\rho_o$ and $a_o$ are the matter density and cosmic scale factor at
maximum expansion.
Obviously, as $t\rightarrow 0$ and $ {2/\sqrt{2\pi G\rho_o}} $,
$a(t)\rightarrow 0$; consequently, $T,{}\rho_m,$ and $R$ all diverge
as expected, since radiation has no effect on spacetime in the former case.

For weakly interacting mixtures of matter and radiation,
recall that these obey (19), (20) and the WEC. Hence they
must sat\-is\-fy $-2<\alpha<-1$ in (22). Since
$_2F_1({1\over 2},{1\over \alpha+2},{\alpha+3 \over \alpha+2}; 0)$
vanishes, it is clear that
this kind of weakly interacting mixture yields a big bang. For example,
when $\alpha=-3/2$, (22) yields
$$
t^2=\bigl(\tau-{8(\sqrt{K_2})^3\over 3\sqrt {K_1}} \bigr)^2
= {16\over 9K_1}(2K_2+ \sqrt a)^2 (K_2-\sqrt a), \eqno(30)
$$
where $\tau$ is the translated proper time.
One can see that the scale factor vanishes at $\tau=0$
and at $\tau=16(\sqrt{K_2})^3/3\sqrt{K_1}$ (see fig.(1)),
yielding respectively a big bang and a big crunch.
In addition, the temperature $T\propto a^{-3/4}$
$\rho \propto a^{-3/2}$ diverge. One can further study the
dynamics of the scale factor $a(t)$ if the assumption of
``weak interactions'' is relaxed; that is, one assumes
the fluids interact significantly and thus the total
energy-momentum tensor will not become the perfect fluid form.

In the case of the FRW models particle horizons will occur if the integral
$$
I=a(t_1)\int_{t_o}^{t_1} {dt\over a(t)} \eqno(31)
$$
converges as $t_o\rightarrow 0$ (big bang) or $-\infty$ (big bang free), where
$t_o$ and $t_1$ are the time of emission of light and time of
reception respectively. $I$ is the proper
distance of the source of emission at $t_1$.
For a $(1+1)$-dimensional
universe with \hbox{non-interacting} mixture of radiation and
matter, we have (29a). It is not difficult to see that
the integral
$$
\int {dt\over a(t)} ={1\over A}ln\bigl({4\pi G\rho_{mo}t \over 2A-4\pi G
\rho_{mo}t}\bigr)+constants\eqno(32)
$$
diverges as one approaches the big bang singularity, $t_o\rightarrow 0$. In
a pure matter ($\rho_r=0$) universe with the solution (29b), we obtain a
similar integral as in (32). Thus particle horizons do not exist in these
two cases. A pure radiation universe yields a cosmic scale factor
$a\propto t$ (or constant). The absence of a big bang in such a universe
implies that the
integral $[\int dt/a(t)]=log_et+constant$ is undefined as $t\rightarrow
-\infty$
and thus no particle horizons exist. This effect is expected since
a pure radiation universe is globally flat.
Therefore a $(1+1)$-dimensional FRW universe with a
\hbox{non-interacting} matter/radiation mixture or pure radiation yields no
particle
horizons.

This former effect is easily seen to be a consequence of the lower
dimensionality of the spacetime. Consider, for example, the
matter-dominated scenario ($\rho_r=0$). For $a(t)\propto t^n$
only those models which have $n<1$ posses a particle
horizon, since (31) implies $I=t_1/(1-n)$ [9].
The \hbox{time-time} component of the \hbox{$(3+1)$-dimensional} Einstein
field equation gives $(d^2a/dt^2)=-(4\pi G/3)\rho a$ with $p=0$. The
dimensionality of our spacetime (four) implies $\rho \propto a^{-3}$
and so $a^2(d^2a/dt^2)=-4\pi G/3$. It is easy to see that in this case
$a\propto t^{2/3}$ thus $n<1$. That is, particle horizons exist.
However in $1+1$ dimensions, the dimensionality of the
spacetime (two) implies $a\propto \rho^{-1}$
implying $n>1$, {\it i.e.} particle horizons do not occur.

\vskip 0.3 true cm

\noindent{\bf 4 Decoupling and Inflation}

In this section, we first discuss an interesting
process called decoupling. Consider a $(1+1)$-dimensional
universe that contains only pure radiation ($\rho_m=0$).
and has a constant cosmic scale factor.
Consequently the temperature and density are also constant.
There is no ``dynamics'' at all. That is, it
is a ``\hbox{steady-state}'' universe without a beginning and an
end. The stress-energy tensor for all matter is traceless.

Suppose that there is some physical mechanism in which the stress-energy
tensor from some form(s) of matter develops a non-zero trace
({\it i.e.} which `decouples' some matter from the radiation).
For simplicity, assume that the change is instantaneous
($\rho_r$ abruptly decreases). Conservation of energy implies
$$
\rho_r = \rho'_r+\rho'_m
$$
where the prime denotes quantities after decoupling. We also assume $a(t)$
and $da/dt$ are continuous at the moment of decoupling, which we take to be
$t_o$. From the viewpoint of a $(1+1)$-dimensional observer, this time
marks the birth of the matter and radiation filled universe. After a time
$t_d-t_o$ this universe will end in a big crunch (see fig.2), where
$t_d=A/2\pi G\rho'_m K_2^2$, where $K_2$ and $A$ are
defined respectively in (22) and (29a).

However, one might argue that the pure radiation universe is ``absolutely
static'': $\rho_r$ is always constant and spacetime does not evolve. How
could some of the radiation spontaneously convert into matter
and what mechanisms provide the conversion? These mechanisms cannot come
from the outside universe since by definition there is no physical system
outside the universe; there is no way it can be ``perturbed''.
Decoupling, then, can only occur if
there are some kinds of macroscopic $(1+1)$-dimensional
quantum fluctuation processes to provide the mechanisms,
the conformal anomaly being the most obvious candidate. However one is then
left with the puzzle of the choice of $t_0$, since the mechanism of the
conformal anomaly should be operative at all times. It would be interesting
to see if a two-dimensional universe could in fact be born via such a
mechanism.

We turn next to a consideration of inflationary processes.
The original form of the inflationary universe
model in $3+1$ dimensions [10] was developed to address the flatness and
horizon problems, (and subsequently the monopole problem).
Although these two cosmological
problems are not present in $1+1$ dimensions, it is instructive to
investigate under what circumstances inflation can occur.

For simplicity, we assume that there exists
a classical scalar field $\phi $ with
an effective potential $V(\phi)$ such that the false
vacuum of $\phi$ has a constant energy density $\rho_f=V(\phi_{false})$,
similar to the \hbox{$(3+1)$-dimensional} case (see e.g. [10]).
$T_{ab}$ then has the perfect fluid form [11]
$$
T_{ab}=-g_{ab}V(\phi_{false})=-g_{ab}\rho_f \eqno(33)
$$
which shows that $p=-\rho_f$. The pressure of the false vacuum
is negative and constant.

Now one can assume that a $(1+1)$-dimensional
homogeneous and isotropic matter and radiation filled
universe begins from a hot big bang and is in a
symmetric phase and then gradually approaches the false
vacuum state (the broken symmetry) as the temperature
drops. The total \hbox{energy-momentum} tensor of the scalar
field becomes equal to (33). To see what happens next,
it is easiest to use (14) and (33) to get
$$
{d^2 a\over dt^2}=-K^2a \eqno(34)
$$
where $K^2=8\pi G\rho_f$ is constant. This equation has the solution
$$
a\propto cos(K(t-t_0)) \eqno(35)
$$
which is not inflationary at all. The reason is that the
field equation (1) in $1+1$ dimensions implies that a
negative pressure tends to slow down the expansion rate
of the universe [$(d^2a/dt^2)<0$]. In $3+1$ dimensions, however,
a negative pressure tends to provide an acceleration
[$(d^2a/dt^2)>0$]. Therefore we conclude that there is no
inflation in $1+1$ dimensions analogous to a $(3+1)$-
dimensional FRW universe under the influence of scalar
fields with positive false vacuum energy density.

The only mechanisms under which exponential inflation can occur involve
adopting unconventional assumptions. Essentially one needs a mechanism
under which the constant $K^2$ on the right hand side of (34) becomes
negative. For example one could consider taking the false vacuum energy
density to be negative. The possibility of the existence of such a negative
energy density depends on particle theories in $1+1$ dimensions which are
beyond the scope of this paper. Another possibility is that the matter
density violates the WEC, that is, $\rho_{mo}<0$ but $\rho_{ro}>0$. On
using (20), we see that $\alpha<-2$ or $\alpha>-1$ (recall that
$\beta=\alpha$ for a non-constant $a(t)$). Now, if $\alpha=0$ ($\rho_m=-
\vert\rho_{mo}\vert=-2\rho_{ro}$), the solution to (21) is
$$
t=\int {da\over \sqrt{K_1} a},\qquad K_1=4\pi G\vert\rho_{mo}\vert \eqno(36)
$$
where $K_2$ (the integration constant) is set to be zero. This integral
obviously
yields
$$
a(t)=a_iexp[\sqrt{K_1}(t-t_i)],\qquad a_i\equiv a(t_i).\eqno(37)
$$
Thus the universe can exponentially inflate under the influence of exotic
matter. During inflation, the energy density of the \hbox{pressure-free}
matter is negative and constant.
Further mechanisms involve a time-dependent constant of gravity
which permits $G$ to become negative or taking $\gamma>2$ in the equation
of state (implying that the speed of sound becomes greater than the speed
of light).

None of the above mechanisms is particularly attractive, and furthermore,
none is necessary. The field equation (1) simultaneously implies that
particle horizons in some cases ({\it e.g.} pure matter or a non-interacting
mixture)
and inflation (under the influence of a positive false vacuum density
scalar field) do not occur in a $(1+1)$-dimensional universe. It would
appear that the lack of structure in the lower-dimensional universe
banishes both the standard problems of $(3+1)$-dimensional cosmology and
the mechanisms which could solve them.

\vskip 0.3 true cm

\noindent{\bf 5 Singularities in FRW and Tilted Models}

In $3+1$ dimensions, spatially homogeneous cosmological
models in general have spacetime singularities [12]. Specifically,
when one considers the spatially homogeneous tilted model, that is,
the fluid velocity \hbox{four-vector} is no longer everywhere and everytime
orthogonal to the homogeneous surfaces, one has either a infinite density
singularity in which $\rho$ diverges or a finite density singularity
in which the tilted angle $\beta$ is unbounded [13]. In section 3
we have shown that infinite density big bang singularities indeed
occur in certain kinds of FRW universes with the equation of
state $p=(\gamma-1)\rho$ .

In this section, without assuming any particular equation of
state, we shall show that infinite density singularities can occur in all FRW
perfect fluid universes (except the pure radiation model) if certain energy
conditions are satisfied. We extend these results to $(1+1)$-dimensional
spatially homogeneous tilted models, showing that
both finite and infinite density singularities can also
happen in this context. We first present the basic equations
and then show that the aforementioned physical
singularities may occur in both \hbox{untilted} and tilted
cases.

As with the $(3+1)$-dimensional tilted model [14], we
may define the quantities $\beta$, $n_a$, $\tilde c_a$ and $c_a$ by the
following relations in $1+1$ dimensions using
$\tilde h_{ab}=g_{ab}+n_a n_b$,
$$
cosh\beta\equiv-u^a n_a,\qquad \tilde h^a{}_b u^b=(sinh\beta)\tilde c^a,
\qquad h^a{}_b n^b=-(sinh\beta)c^a,\eqno(38)
$$
where $\tilde c_an^a=c_au^a=0$, $c^ac_a=\tilde c^a\tilde c_a=1$ and
$\beta(t)$ is the hyperbolic angle of tilt ($t$ is the proper time of the
fluid). As before, $h_{ab}=g_{ab}+u_au_b$.
$n^a$ ($n^an_a=-1$) is a geodesic vector field and normal to $L_t$,
the lines of homogeneity.
$c^a$ is the direction of $n^a$ normal to $u^a$. $\tilde c^a$ is the
direction of the projection of $u^a$ in the lines $L_t$. Since $n_a$ is the
normal geodesic vector field, we can write
$$
n_a=-\tilde t,_{a}, \eqno(39)
$$
and consequently,
$$
\nabla_b n_a = \tilde {\theta}_{ab}, \qquad \tilde {\theta}_{ab}n^b=0.
\eqno(40)
$$
In addition, one can decompose $u^a$ and $n^a$ as
$$
u^a=(sinh\beta)\tilde c^a + (cosh\beta)n^a , \qquad
n^a=-(sinh\beta)c^a+(cosh\beta)u^a.\eqno(41)
$$
By virtue of (41), one can write [14]
$$
d\tilde t = (cosh\beta) dt. \eqno(42)
$$
The first covariant derivative of $u^a$ is given by
$$
\nabla_b u_a=-({d\beta\over d\tilde t})c_a n_b
+ (cosh\beta) \tilde \theta_{ab}
+ (sinh\beta) \nabla_b \tilde c_a, \eqno(43)
$$
on using (41). Consequently, the acceleration and expansion of the fluid
$\theta \equiv \nabla_a u^a$ are respectively given by
$$
{du_a\over dt}={d(sinh\beta) \over d\tilde t}c_a
+ (sinh\beta cosh\beta) \tilde \theta_{ab} \tilde c^b
+ sinh\beta {d\tilde c_a\over dt}, \eqno(44)
$$
and
$$
\theta={d(cosh\beta) \over d\tilde t} +
(cosh\beta) \tilde \theta + (sinh\beta) \nabla_a \tilde c^a. \eqno(45)
$$
In $1+1$ dimensions (45) can be further simplified as follows:
if one constructs normalized {co-ordinates} [$\tilde t$, $\tilde x$],
comoving with the geodesic normals, then in these co-ordinate
$n^a=\delta^a{}_0$, $\tilde h_{a0}=0$, $\tilde h_{11}=g_{11}$. Furthermore,
$$
ds^2=-d\tilde t^2 + g_{11}(\tilde t) d\tilde x^2. \eqno(46)
$$
(In $3+1$ dimensions, in general, $ds^2=-d\tilde t^2
+ g_{\mu \nu}(\tilde t)d\tilde x^{\mu} d\tilde x^{\nu}$, $\mu$, $\nu=1,2,3$,
see e.g. [15]). Using (46)
it is easy to see that $\nabla_0 \tilde c^0 = \nabla_1 \tilde c^1 = 0$
since $\tilde c^a=(0, \tilde c)$ and $\tilde c =\tilde c(\tilde t)$.
This form of $\tilde c^a$ with respect to the metric (46) is obvious since
$n^a \tilde c_a = 0$ and $\tilde c^a$ which spans $L_t$, is invariant
under the group of isometry (one-dimensional spatial translations)
on $L_t$ and therefore it depends only on the ``tilde time''.
Consequently we have in $1+1$ dimensions
$$
\nabla_a \tilde c^a = 0, \eqno(47)
$$
which simplifies (45).

For a perfect fluid one obtains from the components $u_a\nabla_b T^{ab}=0$
and $h^c{}_a\nabla_bT^{ab}=0$ of the conservation equation (15)
$$
(cosh\beta){dlnw\over d\tilde t}+\theta=0,\
\qquad (sinh\beta){dlnr\over d\tilde t}c^a+{du^a\over dt}=0,\eqno(48)
$$
where $w(t)\equiv exp[\int {d\rho\over(\rho+p)}]$ and
$r(t)\equiv exp[\int{dp\over(\rho+p)}]$.
Another expression for acceleration may be obtained from (48):
$$
{du^a\over dt}=(tanh\beta){dp\over d\rho}\theta c^a.\eqno(49)
$$
This equation shows that $du^a/dt$ is parallel to $c^a$ and
vanishes when pressure is constant (or zero).
Finally, combining (44), the second equation of (48) and the
equation $\tilde c^a d\tilde c_a/dt=0$, one has
$$
{dln(rsinh\beta) \over d\tilde t}
+ \tilde \theta_{ab} \tilde c^a \tilde c^b =0. \eqno(50)
$$
This equation shows that $\beta$ is either zero or \hbox{non-zero} for all
$t$.

In $3+1$ dimensions, $\sigma_{ab}=0 \Rightarrow \beta =0$ [14].
Consequently, the only \hbox{shear-free} spatially homogeneous perfect
fluid universe models are the FRW models. A detailed proof involving
($\mu$, $\nu$) and ($0$, $\nu$) field equations and the Jacobi identities
can be found in ref. [14]. In $1+1$ dimensions, however, both
spatial shear and rotation vanish. We must have an FRW universe,
which is not tilted if $du_a/dt=0$. In this case, if $du_a/dt \not= 0$
then $\beta\not= 0$ from (49) and (50) implies that
a tilted model stays tilted. We shall exclude the possibilty that
a universe can have $du_a/dt \not= 0$ and $\beta=0$
at the same time (for a spatially inhomogeneous universe, this situation
is possible).

Conversely, since $(du_a/dt)=0$ implies from (49) that $\beta =0$
(FRW universes cannot be tilted), the following theorem is obvious.

\vskip 0.2 true cm

\itemitem{\bf Theorem 1}: $\beta=0 \Leftrightarrow du_a/dt = 0$ {\sl for
all spatially homogeneous univese models in $1+1$ dimensions}.

\vskip 0.2 true cm

\noindent In other words, the necessary and sufficient condition for a $1+1$
dimensional universe to be a FRW model is that $\beta=0$. Theorem 1
is clearly not true in $3+1$ dimensions. Note also that in $3+1$ dimensions,
in a strict mathematical sense,
spatially isotropic universe models imply spatially homogeneous untilted
models (the converse is not true). It is obvious that in $1+1$ dimensions,
the only \hbox{untilted} models are FRW models.
Hence one gets the following simple theorem:

\vskip 0.2 true cm

\itemitem{\bf Theorem 2}: {\sl For an untilted $(1+1)$-dimensional universe,
it is spatially homogeneous if and only if it is spatially isotropic}.

\vskip 0.2 true cm

Now, rather than assuming any particular equation of state, we show that
physical singularities may happen in both FRW and tilted models under
certain assumptions. Our approach is quite similar to that of Collins
and Ellis [13]. The spacetime we consider will satisfy the following criteria.

\vskip 0.2 true cm

\itemitem{\bf Assumption 1}: ($M$, $g_{ab}$) {\sl is a connected
$(1+1)$-dimensional  $C^{\infty}$ Hausdorff manifold $M$ with a at least $C^2$
Lorentz metric $g$ and it is inextendible and hole-free}.

\vskip 0.2 true cm

\noindent In a strong intuitive sense, this assumption avoids the situation
in which spacetimes which are otherwise \hbox{non-singular} but simply have
points
``artificially removed'' would be considered singular.
In addition, we have assumed that ($M$, $g_{ab}$) has been
extended as far as possible.

Similar to equation (2.2) of Ellis and King [16], we have the following
assumption.

\vskip 0.2 true cm

\itemitem{\bf Assumption 2}: {\sl A $C^1$ equation of state $p=p(\rho)$ for the
fluid is given, and is such that the following conditions}
$$
1\geq dp/d\rho \geq 0,\quad  \rho\geq 0, \quad p\geq 0, \quad \rho\geq p
$$
{\sl hold at all times}.

\vskip 0.2 true cm

\noindent The first inequality expresses the fact that the speed of sound
cannot exceed unity ({\it i.e.} the speed of light) in $1+1$ dimensions,
and that the initial value problem is well posed [17]. The second and
fourth ones are the WEC and SEC (strong energy condition) respectively.
The third one is imposed since we are not going to involve a false vacuum
(or equivalently, a cosmological constant) in our dicussion. Assumption 2
implies that there is a time $t_o$ such that $\rho_o+p_o>0$, where $\rho_o
=\rho (t_o)$ and $p_o=p(t_o)$.

The average length scale, ${\cal R}(t)$, is defined as $(d{\cal R}/dt)(1/{\cal
R})=\theta$
(${\cal R}(t)=a(t)$ in FRW models). We assume that the universe is expanding
at the time $t_o$.

\vskip 0.2 true cm

\itemitem{\bf Assumption 3}: {\sl On each world line there is a time $t_o$
such that $(d{\cal R}/dt)_o(1/{\cal R})_o \equiv H_o>0$ }.

\vskip 0.2 true cm

We first consider FRW models. Recall that in such models, (14) and (16)
are the dynamical equations. In addition, because of (16), one can
integrate (14) to obtain the Friedmann equation in $1+1$ dimensions,
$$
({d{\cal R}\over dt})^2=-8\pi G{\cal R}^2\rho + C, \eqno(51)
$$
where C is a constant. Now (16) and the first inequality show that
$$
0<{\cal R}\leq {\cal R}_o \Rightarrow (\rho_o+p_o)({{\cal R}_o\over {\cal
R}})^2 \geq (\rho+p)
\geq (\rho_o+p_o)({{\cal R}_o\over {\cal R}}) \geq \rho_o+p_o >0. \eqno(52)
$$
The mean value theorem of a $C^1$ curve with the first inequality of
assumption 2 implies that $1\geq (p_2-p_1)/(\rho_2-\rho_1) \geq 0 $ for the
$C^1$ curve $p=(\rho)$. However we have just shown that $(\rho +p)\geq
(\rho_o +p_o)$. So, it must be true that
$$
p\geq p_o, \qquad \rho\geq \rho_0.\eqno(53)
$$
Conditions (52), (53) and the mean value theorem yield
$$
(\rho_o +p_o)({{\cal R}_o^2\over {\cal R}^2}-1)+\rho_o -p_o \geq \rho -p \geq
0, \eqno(54)
$$
and
$$
(\rho_o +p_o)({{\cal R}_o^2\over {\cal R}^2}-1)+\rho_o \geq \rho \geq {1\over
2}(\rho_o +p_o)({{\cal R}_o\over {\cal R}}-1)+\rho_o. \eqno(55)
$$
Note that the term $\rho -p$ does not have the similar expression for
its lower bound as $\rho +p$ and $\rho$. In fact, we have used the SEC
rather than (52) and (53) (a point to which we shall later return). From
(52), (54), (55), (14) and (51), it is clear that:

\vskip 0.2 true cm

\vbox{{\sl For each value of ${\cal R}$ with $0<{\cal R}<{\cal R}_o$, ${1\over
{\cal R}}{d^2{\cal R}\over dt^2}$
and ${1\over {\cal R}}{d{\cal R}\over dt}$ are bounded{\hfill}
above and below by finite bounds, as are
$\rho$, $\rho +p$ and $\rho-p$}}.{\hfill (56)}

\vskip 0.2 true cm

\noindent In addition, (14) and the SEC show that
$$
0<{\cal R}\leq {\cal R}_o \Rightarrow ({d^2{\cal R}\over dt^2})<0. \eqno(57)
$$
Let the universe be regular and well-defined for $T_1<t\leq t_o$.
Now condition (57) and assumption 3 show that
$$
T_1<t \leq t_o \Rightarrow 0<{\cal R}\leq {\cal R}_o \Rightarrow
{d{\cal R}\over dt}\geq ({d{\cal R}\over dt})_o = H_o{\cal R}_o>0. \eqno(58)
$$
Thus
$$
0<{\cal R}\leq {\cal R}_o(1-H_o(t_o-t)),\eqno(59)
$$
for $T_1<t\leq t_o$. Obviously, ${\cal R}(t)$ is a monotonically decreasing
function if one goes backward in time from $t_o$.
The geometry of the universe is regular when ${\cal R}>0$ due
to condition (56), and thus the initial data,
[$\rho$, ${\cal R}$, $d{\cal R}/dt$], for the (14) and (16) on a spacelike line
(constant $t$) is well-behaved. In addition,
recall that we have assumed that initial value problem is well posed in
$1+1$ dimensions (assumption 2). Consequently, we do not expect some
other singularity to
intervene at a time $T_1$ or earlier times when ${\cal R}$ is still bounded
away
from zero. Assumption 1 states that if it is possible to
extend the universe, it is so extended. Hence, (59) implies
that ${\cal R}\rightarrow 0$ a finite time ago.
Without loss of generality, we take
${\cal R}\rightarrow 0$ as $t\rightarrow 0^+$. Now incorporating conditions
(52)
and (55), the following conclusion can be drawn:

\vskip 0.2 true cm

\itemitem{}All $(1+1)$-dimensional FRW universe models satisfying
assumptions 1-3, and (14) and (16) must have physical singularities,
that is, as $t\rightarrow 0^+$, ${\cal R}\rightarrow 0$, $\rho\rightarrow
\infty$,
$\rho +p\rightarrow \infty$ and
$R_{ab}R^{ab}\propto (p-\rho)^2 \rightarrow\infty$.

\vskip 0.2 true cm

\noindent Note that both $p$ and $\rho$ diverge as ${\cal R}\rightarrow 0$,
and thus it is possible that $p-\rho$ (along with $R_{ab}R^{ab}$) is finite.
Unless $p=\rho$, we exclude any equations of state in which
the divergence in $p$ ``cancels'' the divergence in $\rho$ and makes
$R_{ab}R^{ab}$ finite.

It is clear that this conclusion does not
hold for a universe of pure radiation since it does not satisfy assumption 3.
In addition, recall that zero was taken to be the lower
bound for the term $(\rho-p)$ in (54).
If it is possible to find a lower bound for $\rho-p$
similar to the forms of (52) and (55), then
$\rho_r -p_r$ will diverge as ${\cal R}\rightarrow 0$. However, in the special
case
of a non-interacting mixture with equation of state $p=(\gamma-1)\rho$ , this
obviously
cannot be true since one always has $\rho_r -p_r =0$. It is due
to the fact that for this equation of state and field
equation (1), radiation has no effect on the
structure of $(1+1)$-dimensional spacetimes.

Finally, by considering the family of geodesic curves normal to $L_t$, the
family of lines of homogeneity, we show that
infinite density or ``whimper'' singularities
may happen in $(1+1)$-dimensional tilted models. The reason we do not
consider the family of fluid flow lines is that $du_a/dt\not=0$
and we do not know directly how $\nabla_a(du^a/dt)$
varies with ${\cal R}(t)$. On the other hand, for the geodesic normals the
acceleration vanishes so that we can argue as before.
We define the average
length scale for the normal geodesic as $(d{\tilde {\cal R}}/d{\tilde
t})(1/\tilde
{\cal R})=\tilde \theta$. Relative to the vector field $n^a$, the Ricci tensor
takes the form
$$
R_{ab}={1\over 2}g_{ab}R=4\pi Gg_{ab}(\tilde p-\tilde\rho) \eqno(60)
$$
where $T=p-\rho=\tilde p -\tilde\rho$. Here $\tilde\rho =
\rho(cosh^{2}\beta) + p(sinh^{2}\beta)$ and $\tilde p = p+(\rho +
p)(sinh^{2}\beta)$ are the energy density and pressure respectively
measured in the normal geodesic frame (see (1.33b) of [14]). Thus the
Raychaudhuri equation for $\tilde \theta$ takes exactly the same form as
(13) or (14) if $\tilde {\cal R}(\tilde t)$ is being used instead. Because
of the similarity to the previous equations,
much of the argument is parallel to
the previous case. Then the analogues of (57), (58) and (59) show that
$\tilde {\cal R} \rightarrow 0$ a finite time ago (as before, we shall take
$\tilde t\rightarrow 0^+$). Note that relative to $n^a$, the
\hbox{energy-momentum} tensor is not that of a
perfect fluid and therefore (15) will not reduce to (16).

In order to get the analogue of condition (55), we use the quantity $w(t)$,
the fluid enthalpy defined in (48). This definition and assumption 2 yield
$$
\rho \geq \rho_o \Rightarrow {\rho\over \rho_o}\geq {w\over w_o}
\geq ({\rho\over \rho_o})^{1/2}. \eqno (61)
$$
Now (45), (47) and the first equation of (48) give
$$
{d(ln(\tilde {\cal R} wcosh\beta))\over d\tilde t}=0. \eqno(62)
$$
According to this equation, one has
$$
w (cosh\beta) \propto {1\over \tilde {\cal R}}, \eqno(63)
$$
which is a generalization of condition (55).
Since $\tilde t \rightarrow 0^+$, $\tilde {\cal R} \rightarrow 0$
we conclude that $\lim_{t\to 0^+}\rho (cosh\beta) \rightarrow \infty$.
It is easy to see that one either has
$$
\tilde t\rightarrow 0^+\Rightarrow \rho \rightarrow \infty,
\qquad g^{ab}R_{ab} \rightarrow \infty, \eqno(64)
$$
or
$$
\tilde t\rightarrow 0^+\Rightarrow \beta \rightarrow \infty \eqno(65)
$$
with $\rho$ bounded. From the normal geodesic viewpoint,
(64) corresponds to an infinite density singularity.
On the other hand, (65) does not give a diverging Ricci
scalar. An observer moving on these geodesics will
experience a tidal force determined by the
geodesic deviation equation in $1+1$ dimension,
$$
R_{abcd}X^cn^bn^d=
(1/2)g^{ef}R_{ef}(-g_{ac}-g_{ad}n^dg_{bc}n^b)X^c
=4\pi G(\rho-p)X_a, \eqno (66)
$$
where $X^a$ is the normal deviation vector from a geodesic to
an infinitesimally nearby geodesic. In some chosen frames if $g_{ab}$
blows up as $\tilde t\rightarrow 0^+$, the tidal force will
blow up as well. Looked at from the fluid viewpoint, similar
events happen. One also has a infinite density singularity
in (64). For the situation in (65), one can
choose $[n^a, u^a]$ as a non-orthogonal basis
for the ``tilted spacetime'', we have, in this basis,
$g_{nu}=n^au_a=-cosh\beta$ which blows up when
$\tilde t\rightarrow 0^+$. As (42) implies $\tilde dt>dt$, there
is a finite value $T_1$ such that when $\tilde t\rightarrow 0^+$
on these fluid flow lines, $t\rightarrow T^+_1$. Of course,
the bad behaviour of the metric component may be a
co-ordinate singularity rather than a physical one.
Hence, as $t\rightarrow T^+_1$ along the fluid flow lines,
either the flow lines may run into a finite density singularity
or it is possible to extend the flow lines beyond $t=T_1$.
In this fashion a ``frame-dependent whimper'' singularity
can occur in (65).

\vskip 0.3 true cm

\noindent{\bf Conclusions}

Our exploration of $(1+1)$ dimensional cosmologies using (1) has shown that
such cosmologies for the most part bear a strong qualitative resemblance to
their $(3+1)$ dimensional counterparts. The most notable departure from
this is, perhaps, the absence of both particle horizons and the possibility
of realizing an inflationary scenario without violating the WEC or imposing
some other unconventional condition. The most significant resemblance is
the persistent occurance of singularities for virtually all models except
the radiation dominated scenario.  Whether or not quantum gravitational and
quantum cosmological effects will be able to appropriately treat such
singularities is a subject for further study.

\vskip 0.3 true cm

\noindent{\bf Acknowledgements}

\noindent This work was supported by the Natural Sciences and Engineering
Research Council of Canada. We would like to thank G. Ellis for
discussions.

\vskip 0.3 true cm

\noindent{\bf Appendix}

In this appendix we give a toy model for inflationary scenario
in two spacetime dimensions.

Suppose the life history of a $(1+1)$-dimensional
universe is divided into three epochs according to one of the two following
scenarios:

\vskip 0.2 true cm

\itemitem{\bf 1} Big Bang (non-interacting matter + radiation)
$\longrightarrow$
Inflation (exotic matter and radiation) $\longrightarrow$
(non-interacting matter + radiation)
\itemitem{\bf 2} Big Bang (non-interacting matter + radiation)
$\longrightarrow$
Inflation (exotic matter and radiation) $\longrightarrow$ radiation only.

\vskip 0.2 true cm

\noindent The second sequence is trivial since the universe suddenly
becomes static. Note that just before inflation, all matter spontaneously
and instantly converts into exotic matter and consequently this exotic
matter ``accelerates'' the expansion. As discussed in section 4 some
physical mechanism operative at macroscopic distance scales in $1+1$
dimensions is required to carry out such a conversion; a scalar field with
negative energy density false vacuum would work, for example.

Assuming that such a mechanism is possible, we continue to study
sequence {\bf 1}. At each transition between the two epochs, both $a(t)$ and
$da/dt$ are assumed to be continuous.

For epoch 1 (the matter and radiation era, $0\leq t\leq t_i$), we
choose $t=0$ as the time when $a(t)=0$ (origin of the universe). Then
$$
a(t)=-Bt^2+At,\qquad B=2\pi G\rho_{mo},\eqno(A1)
$$
$$
{da\over dt}=-2Bt+A.\eqno(A2)
$$
For epoch 2 (the inflationary era, $t_i\leq t\leq t_f$),
all the matter instantly becomes exotic and the negative
energy density is constant during inflation:
$$
a=a_iexp[\sqrt{K_1}(t-t_i)], \qquad a_i\equiv a(t_i)\> of\> (A1),\eqno(A3)
$$
$$
{da \over dt}=a_i\sqrt{K_1}exp[\sqrt{K_1}(t-t_i)].\eqno(A4)
$$
Note that $a_iK_1=2B$ which expresses the fact that all ordinary matter
becomes exotic. The continuity of $a(t)$ and $da/dt$ at $t_i$ implies
$$
t_i^{\pm}=(1/2 \pm \sqrt{3}/6)(A/B). \eqno(A5)
$$
There are two solutions for $t_i$. However, $t_i^+>t=A/2B$, the time
when the universe starts to contract; inflation can only start at $t_i^-$.
Knowledge of A and B (or $\rho_{mo}$), permits one to determine
$t_i^-$. In other words, the ``initial conditions'' $A$ and $B$ exactly
determine the time when the ordinary matter will becomes exotic.

For epoch 3 all the exotic matter instantly convert back to ordinary matter.
For simplicity we assume $a(0)=0$. For the interval $t_f\leq t\leq t_o$:
$$
a(t)=-B't^2+A't,\qquad B'=2\pi G\rho_{mo}'. \eqno(A6)
$$
Note that, $\rho_{mo}=({a_i/ a_f})\rho_{mo}'$. Thus $a_fK_1=2B'$.
Similarly, the continuity of $a(t)$ and $da/dt$ at $t_f$ implies
$$
t_f^-=(1/2 - \sqrt{3}/6)(A'/B'). \eqno(A7)
$$
$t_f^+$ is omitted for the same reason as above.
Since $B'=B(a_f/a_i)$ and $t_f^->t_i^-$, we must have
$A'>A(a_f/a_i)$. Knowledge of $A'$ and $\rho_{mo}'$ permits
determination of $t_f^-$, that is, the time when all the exotic matter
become ordinary again.

\vskip 0.6 true cm

{\bf References}

\item{1.} C. Teitelboim in {\sl Quantum Theory of Gravity},
ed. S. Christensen (Adam Hilger, Bristol 1984), p.327; R. Jackiw,
{\it ibid}., p.~403;  Nucl. Phys. {\bf B252} (1985) 343;
C. Teitelboim, Phys. Lett. {\bf B126} (1983) 41,46;
J. Gegenberg, P.F. Kelly, R.B. Mann, and D.E.~Vincent,
Phys. Rev. {\bf D37} (1988) 3463;
P.F. Kelly and R.B. Mann Phys. Rev. {\bf D43} (1991) 1839;
J. Polchinski, Nucl. Phys. {\bf B324} (1989) 123; J. Harvey and A.
Strominger, `Quantum Theory of Black Holes' EFI-41 (1992).
\item{2.} R.B. Mann, Found. Phys. Lett. {\bf 4} (1991) 425; Gen. Rel.
Grav. {\bf 24} (1992) 433.
\item{3.} R.B. Mann, S. Morsink, A.E. Sikkema and T.G.
Steele, Phys. Rev. {\bf D43} (1991) 3948.
\item{4.} R.B. Mann and S.F. Ross,
University of Waterloo preprint, WATPHYS-TH92/06.
\item{5.} R.B. Mann, A. Shiekh, and L. Tarasov, Nucl. Phys. {\bf B341}
(1990) 134; A.E. Sikkema and R.B. Mann, Class. Quantum Grav.
{\bf 8} (1991) 219; R.B. Mann and T.G. Steele, Class. Quantum Grav. {\bf 9}
(1992) 475; S.M. Morsink and R.B. Mann, Class. Quant. Grav. {\bf 8} (1991)
2215; D. Christensen and R.B. Mann, Class. Quant. Grav. {\bf 9} (1992)
1769.
\item{6.} G.F.R. Ellis, {\sl Cargese Lectures in Physics}, vol. 6, E.
Schatzmann, ed. (Gordon and Beach, New York, 1973).
\item{7.} R. Wald, {\sl General Relativity}
(University of Chicago Press, Chicago, 1984).
\item{8.} F.C. Andrews,
{\sl Equilibrium Statistical Mechanics} (Wiley, New York, 1975).
\item{9.} W. Rindler,  Mon. Not. R. Astron. Soc. {\bf 116} (1956) 662.
\item{10.} A. Guth, Phys. Rev. {\bf D23} (1981) 347.
\item{11.} G.F.R. Ellis, in {\sl Gravitation: a Banff Summer Institute},
eds. R.B. Mann, and P.S. Wesson, (World Scientific, Singapore, 1991).
\item{12.} S.W. Hawking and G.F.R. Ellis, {\sl The large scale
structure of Spacetime} (Cambridge University Press, Cambridge, 1973).
\item{13.} C.B. Collins and G.F.R. Ellis, Phys. Rep. {\bf 56} (1979) 65.
\item{14.} A.R. King and G.F.R. Ellis, Commun. Math. Phys. {\bf 31} (1973)
209.
\item{15.} M.A.H. MacCallum in {\sl General Relativity,
an Einstein Centenary Survey}, S.W. Hawking and W. Israel, ed.
(Cambridge University Press, Cambridge, 1979).
\item{16.} G.F.R. Ellis and A.R. King, Commun. Math. Phys. {\bf 38} (1974) 119.
\item{17.} M. Kriele, Class. Quant. Grav. (To be published).

\end

%% file: paper.tex
\magnification=1000
\hsize=6.5 true in
\vsize=8.5 true in
\normalbaselineskip=18pt plus 3pt

\def\doublespace{\baselineskip=18pt plus 7pt\message{double space}}
\def\singlespace{\baselineskip=10pt plus 3pt\message{single space}}
\let\spacing=\doublespace
\parindent=1.0 true cm

\def\body{\vfill\eject\parindent=1.0 true cm
 \ifx\spacing\singlespace\singlespace\else\doublespace\fi}

\def\title#1{\centerline{{\bf #1}}}




\def\ref#1{$^{(#1)}$}    

\def\upwa#1{\raise 1pt\hbox{\sevenrm #1}}
\def\dnwa#1{\lower 1pt\hbox{\sevenrm #1}}
\def\dnsa#1{\lower 1pt\hbox{$\scriptstyle #1$}}
\def\upsa#1{\raise 1pt\hbox{$\scriptstyle #1$}}
\def\upwb#1{\raise 2pt\hbox{\sevenrm #1}}
\def\dnwb#1{\lower 2pt\hbox{\sevenrm #1}}
\def\dnsb#1{\lower 2pt\hbox{$\scriptstyle #1$}}
\def\upsb#1{\raise 2pt\hbox{$\scriptstyle #1$}}
\def\upwc#1{\raise 3pt\hbox{\sevenrm #1}}
\def\dnwc#1{\lower 3pt\hbox{\sevenrm #1}}
\def\upsc#1{\raise 3pt\hbox{$\scriptstyle #1$}}
\def\dnsc#1{\lower 3pt\hbox{$\scriptstyle #1$}}

\def\ccom{\,\raise2pt\hbox{,}} 
\def\hprime{\raise 2pt\hbox{$\scriptstyle \prime$}} 


\catcode`@=11
\def\C@ncel#1#2{\ooalign{$\hfil#1\mkern2mu/\hfil$\crcr$#1#2$}}
\def\gf#1{\mathrel{\mathpalette\c@ncel#1}}      
\def\Gf#1{\mathrel{\mathpalette\C@ncel#1}}      

\def\gapx{\lower 2pt \hbox{$\buildrel>\over{\scriptstyle{\sim}}$}}
\def\lapx{\lower 2pt \hbox{$\buildrel<\over{\scriptstyle{\sim}}$}}




\def\ref#1{$^{(#1)}$}    


\def\ccom{\,\raise2pt\hbox{,}} 

\def\dn#1{\lower 2 pt\hbox{$\scriptstyle #1$}}
\def\up#1#2{\raise {#1} pt\hbox{$\scriptstyle #2$}}
\def\hprime{\raise 2pt\hbox{$\scriptstyle \prime$}}


\def\sec{\hbox{\lower 1pt\rlap{{\sixrm S}}{\hbox{\raise 1pt\hbox{\sixrm S}}}}}


\def\sectionskip{\penalty-500\vskip24pt plus12pt minus6pt}

\def\SECTION#1#2\par{\goodbreak
   \sectionskip\leftline{\bf #1\quad #2}\nobreak\vskip0.3cm
   \mark{#1\quad #2}}

         \newtoks\chaptitle
         \newcount\startpage

         \def\makeheadline{\vbox to 0pt{\vskip -22.5pt
                           \line{\vbox to 8.5pt{}\the\headline}
                           \vss}\vskip 24pt\nointerlineskip}

         \def\optiona{\hfil\rlap{\kern 12 pt\bf\folio}}
         \def\optionb{\hfil}
         \def\start{\ifnum\pageno<0\optionb\else\optiona\fi}

         \headline={\ifnum\pageno=\startpage\start\else\optionb\fi}





\newcount\eno
\def\assigneno #1{\advance\eno by 1 \let\count{#1}=\eno
      \eqno(\hbox{$\eno$})}

%
\def\abstract#1{\vskip 15 pt \midinsert \narrower \smallskip
                \noindent{\bf ABSTRACT.} \quad {#1}
                \smallskip \endinsert}
%
